\documentstyle[prl,aps,preprint,psfig]{revtex}
                        
\begin{document}

\draft
\preprint{TU-593}
\title{%
Neutrino Masses and Mixing from Bilinear R-Parity Violation
\footnote{To appear in conference proceedings "Neutrino Oscillations
  and their Origin", Fuji-Yoshida, Japan, Feb. 11-13, 2000} 
}
\author{Fumihiro Takayama\footnote{e-mail: takayama@hep.phys.tohoku.ac.jp}
 and Masahiro Yamaguchi\footnote{e-mail: yama@hep.phys.tohoku.ac.jp}}
\address{Department of Physics, Tohoku University \\
         Sendai 980-8578, Japan}

\date{May 2000}

\maketitle
\begin{abstract}
Neutrino masses and mixing are generated in a supersymmetric standard
model when R-parity is violated in bilinear mass terms. The mixing
matrix among the neutrinos takes a restrictive form if the
lepton flavor universality holds in the R-parity violating soft masses. It
turns out that only the small angle MSW solution to the solar neutrino
problem is consistent with the result of the CHOOZ experiment and 
the atmospheric neutrino data.
\end{abstract}

\clearpage

\section{Introduction}

R-parity violation in a supersymmetric standard model provides an
intriguing mechanism to generate neutrino masses. The R-parity is a
$Z_2$ discrete parity which distinguishes (R-parity odd)
superparticles from (R-parity even) ordinary particles.  Though one
often assumes the R-parity conservation in model building, it can break
 without conflicting phenomenological problems such as very 
fast proton decays.
If the R-parity violating terms also violate lepton number
conservation, the neutrinos acquire masses either at tree level or at
loop levels \cite{TY-HallSuzuki}. 
One of the appealing points of this scenario is that it
does not require existence of exotic particles such as heavy
right-handed neutrinos apart from the superpartners which are already
included in the minimal supersymmetric standard model.

The R-parity violation may also give novel collider signatures. If the
R-parity were conserved, superparticles would be produced in pairs,
and also the lightest superparticle (LSP) would be stable and escape
detection, resulting in a missing energy as a supersymmetry signal. If
the R-parity is broken, on the other hand, the lightest superparticle
may decay to ordinary particles inside a detector.  Detailed study of
the final states may reveal properties of the R-violating
interaction. Thus in this scenario, useful information on the neutrino
masses may also be inferred  from collider experiments.

Here we would like to consider the case of bilinear R-parity violation
in which R-parity is broken only in bilinear mass terms. This is
particularly interesting because it can be embedded into a grand
unified theory (GUT) where quarks and leptons belong to one and the same
representation of a GUT group.  We will argue that, when the lepton
flavor universality holds in the soft supersymmetry breaking masses,
the neutrino mixing matrix is in a special form which is parameterized only
by two angles. Thus the resulting pattern of the neutrino oscillations should
be restricted. In fact we find that when we combine the CHOOZ experiment
result with the $\nu_{\mu}$-$\nu_{\tau}$ oscillation solution to the
atmospheric neutrino, the only allowed solar neutrino solution is the
small angle MSW \cite{TY-TakayamaYamaguchi}. 

The results presented here are essentially given in 
Ref.~\cite{TY-TakayamaYamaguchi} already, but we slightly generalize the 
previous ones. Namely here we discuss the case where the lepton flavor 
universality among soft supersymmetry breaking masses is assumed, 
whereas in the previous paper 
\cite{TY-TakayamaYamaguchi} the lepton-Higgs universality 
was considered. 
See Ref.~\cite{TY-TakayamaYamaguchi} and references therein for more details.

\section{Bilinear R-parity Violation}

We first explain the model we are considering.  The particle contents of
the model are  those of the minimal supersymmetric standard model
(MSSM). We shall assume R-parity breaking bilinear terms  in superpotential
\begin{eqnarray}
  W = \mu H_D H_U +\mu_i L_i H_U +Y^L_i L_i H_D E^c_i +Y^D_i Q_iH_D D^c_i 
 +Y^U_{ij} Q_i H_U U^c_j. \label{eq:superpotential}  
\end{eqnarray}
Here $H_D$, $H_U$ are two Higgs doublets, $L_i$ a $SU(2)_L$ doublet lepton, 
$E^c_i$ is a singlet lepton, $Q_i$  a doublet quark, $U^c_i$, $D^c_i$ a 
singlet quark of up and down type, respectively. Suffices $i, j$ stand
for generations. 
The soft SUSY breaking terms in the scalar potential are
\begin{eqnarray}
  V^{\mathrm{soft}}& =&B H_D H_U+B_i \tilde{L}_i H_U
 +m_{H_D}^2H_DH_D^{\dagger}+m_{H_U}^2H_UH_U^{\dagger}
\nonumber \\
& &  +m_{HL_i}^2\tilde{L}_iH_D^{\dagger}+m_{L_{ij}}^2
  \tilde{L}_i\tilde{L}_j^{\dagger} +\cdots.
\end{eqnarray}
where we have written only bilinear terms explicitly. 
Here we assume the following lepton flavor universality in the soft masses:
\begin{eqnarray}
B_i \propto \mu_i,~ 
m^2_{L_{ij}} \propto \delta_{ij},~ 
m_{HL_i}^2 \propto \mu_i \label{eq:lH-universality}.
\end{eqnarray}
This universality suffers from radiative
corrections and we assume that Eq.~(\ref{eq:lH-universality}) holds at an
energy scale where these soft masses are given as the boundary conditions
of the renormalization group equations.

In this model, the R-parity violating terms are parameterized by
\begin{eqnarray}
  s_3&\equiv&\sin \theta_3=
\frac{\sqrt{\mu_1^2+\mu_2^2+\mu_3^2}}{\sqrt{\mu_1^2+\mu_2^2+\mu_3^2+\mu^2}},
\nonumber\\
  s_2&\equiv&\sin \theta_2=
      \frac{\sqrt{\mu_1^2+\mu_2^2}}{\sqrt{\mu_1^2+\mu_2^2+\mu_3^2}},
\nonumber\\
  s_1&\equiv&\sin \theta_1=\frac{\mu_1}{\sqrt{ \mu_1^2+\mu_2^2}}
\end{eqnarray}
Here, for simplicity, we have taken $\mu$ and $\mu_i$ to be real.
$s_3$ represents the magnitude of the R-parity violation, while the
other two parameters characterize the mixing of the neutrinos.

\section{MNS Mixing Matrix}
Let us now compute the neutrino masses and their mixing. To do this, it is
convenient to use the Lagrangian whose renormalization point is at the
electroweak scale. This may be obtained by the use of the renormalization
group. Using this technique it is easy to see that one combination
of the sneutrino fields $s_1 s_2 \tilde{\nu}_{e}+c_1s_2 \tilde{\nu}_{\mu} 
+c_2 \tilde{\nu}_{\tau}$ dominantly develops a non-vanishing vacuum 
expectation value (VEV). The VEV induces a mixing between the neutrino and
neutralinos, generating a mass for the neutrino at the tree level.
The neutinos also acquire masses at the one-loop level.  An analysis of 
Ref~\cite{TY-TakayamaYamaguchi} 
shows that the mixing matrix of the neutrino sector, the 
MNS matrix \cite{TY-MNS}, becomes 
\begin{equation} 
  U_{i \alpha}=
\left(
    \begin{array}{@{\,}ccc@{\,}}
      U_{\tau 3} & U_{\tau 2} & U_{\tau 1} \\
      U_{\mu 3}  & U_{\mu 2}  & U_{\mu 1}  \\
      U_{e 3}    & U_{e 2}    & U_{e 1}    \\
    \end{array}
\right) 
=
  \left(
    \begin{array}{@{\,}ccc@{\,}}
      c_{\theta} & -s_{\theta} & 0 \\
      c_1s_{\theta} & c_1c_{\theta} & -s_1 \\
      s_1s_{\theta} & s_1c_{\theta} & c_1 \\
    \end{array}    
  \right),   
\end{equation}
where $\theta$ is approximately equal to $\theta_2$ with some small 
correction. Here
$i$ and $\alpha$ denote the weak current basis and the mass eigen
basis, respectively. The MNS matrix obtained here contains  only two angles,
while a general $3\times 3$ rotation matrix will  have 3 angles. This 
is essential in the following analysis.

\section{Results}

We are now at the position to give our results. 
Here we assume the hierarchical mass structure, namely $m_3 \gg m_2 \gg m_1$
so that $\Delta m_{32}^2 \simeq \Delta m_{31}^2 \gg \Delta m_{21}^2$, which 
are  naturally realized in our model. 

The atmospheric neutrino can be explained by $\nu_{\tau}$-$\nu_{\mu}$ 
oscillation. The transition probability is proportional to 
$4|U_{\mu3}|^2|U_{\tau3}|^2=4c_1^2s_{\theta}^2c_{\theta}^2$ 
and it must be close to unity to accord with the superKamiokande data 
\cite{TY-atm}.  
On the other hand, the CHOOZ experiment \cite{TY-CHOOZ}
gives a bound on
$4|U_{e3}|^2=4s_1^2 s_{\theta}^2$ to be smaller than 0.2. These two 
constraints imply that the angle $s_{1}$ must be small. Therefore the
solar neutrino \cite{TY-solar}
 must be explained by the small angle MSW solution, since
$\nu_{\mu}$-$\nu_{e}$ oscillation involves the small $s_{1}$ in
its transition probability. In fact the large angle solutions require
a large $s_1$, in contradition with the other experimental results.
This non-trivial relation comes from the fact that the MNS matrix is 
characterized by the two angles. 
If one relaxed the lepton flavor universality among the soft masses, one would
get a more general mixing matrix. However, one should carefully
choose the soft masses not to conflict with the severe bounds on
the lepton flavor violating processes.

Next we would like to discuss the neutrino masses. The atmospheric neutrino
requires $\Delta m_{\rm{atm}}^2 \simeq (2-6) \times 10^{-3} \mbox{eV}^2$
and the small angle MSW to the solar neutrino indicates 
$\Delta m^2_{\rm{SMSW}} \simeq (0.4-1)\times  10^{-5} \mbox{eV}^2$.
Since in our scenario the heaviest neutrino mass is obtained at the tree 
level while the next one is generated at the one loop, the neutrino masses tend
to have large hierarchy. The less hierarchical structure suggested by
the experiments requires a mild fine tuning of the tree level VEV of the 
neutrino to suppress the tree-level mass. This can easily be achieved in many
ways, one of which is the universality between the leptons and the Higgs in the
soft masses and the use of the alignment.

\section{Conclusions}

To summarize, we have considered the case of the bilinear R-parity
violation with lepton flavor universality among the soft supersymmetry
breaking masses. This generates the neutrino masses and mixing. The
mixing matrix of the neutrinos has a very special pattern. This leads
us to conclude that the large mixing angle solutions to the solar
neutrino problem are ruled out when the CHOOZ result and the
atmospheric neutrino data are combined together. Furthermore the
relatively less hierarchical structure of the neutrino masses in this
case are obtained if the soft SUSY breaking masses are suitably tuned
to give small VEV for sneutrinos. It is interesting to mention that
neutrino oscillation experiments,
{\it e.g.} SuperKamiokande, SNO\cite{TY-SNO}, and
KamLAND\cite{TY-KamLAND}, as well as collider experiments in future 
will provide (critical) tests to our scenario.

%%%%%%%%%%%%%%%%%%%%%%%%%%%%%%%%%%%%%%%
%%%%%%%%%%%%%%%%%%%%%%%%%%%%%%%%%%%%%%%
%%                                                   %%%%%%%%%%%%%%
%%    thebibliography environment   %%%%%%%%%%%%%%
%%                                                    %%%%%%%%%%%%%%
%%%%%%%%%%%%%%%%%%%%%%%%%%%%%%%%%%%%%%%
%%%%%%%%%%%%%%%%%%%%%%%%%

\end{document}